\documentclass[12pt,preprint]{aastex}

\newcommand{\om}{\ensuremath{\Omega_m}}

\newcommand{\ola}{\ensuremath{\Omega_{\Lambda}}}
\newcommand{\kmsmpc}{{\ensuremath{{\rm km~s}^{-1}~{\rm Mpc}^{-1}}}}

\newcommand{\cosmol}[3]{\ensuremath{\om = #1,\;\ola = #2,\;H_0 = #3~\kmsmpc}}
\newcommand{\lcdmparm}{\cosmol{0.24}{0.76}{73}}
\newcommand{\etal}{et~al.\/}
\newcommand{\lya}{Ly\ensuremath{\alpha}}

\newcommand{\msun}{\ensuremath{M_\sun}}
\newcommand{\dmst}[3]{\ensuremath{{#1}\degr\,{#2}\arcmin\,{#3}\arcsec}}
\newcommand{\hmst}[3]{\ensuremath{{#1}^h\,{#2}^m\,{#3}^s}}
\newcommand{\dms}[3]{\ensuremath{{#1}\ {#2}\ {#3}}}
\newcommand{\hms}[3]{\ensuremath{{#1}\ {#2}\ {#3}}}

\newcommand{\sqdeg}{square degree}
\newcommand{\sqdegs}{square degrees}
\newcommand{\bootes}{Bo\"otes}
\newcommand{\xbootes}{XBo\"otes}
\newcommand{\ujy}{\ensuremath{\mu}Jy}
\newcommand{\uJy}{\ensuremath{\mu}Jy}
\newcommand{\ie}{i.\,e.\,}
\newcommand{\eg}{e.\,g.\,}
\newcommand{\esca}{erg s\ensuremath{^{-1}} cm\ensuremath{^{-2}} \AA\ensuremath{^{-1}}}
\newcommand{\esc}{erg s\ensuremath{^{-1}} cm\ensuremath{^{-2}}}

\newcommand{\whzsr}{W Hz\ensuremath{^{-1}} sr\ensuremath{^{-1}}}
\newcommand{\ks}{\ensuremath{K_s}}
\newcommand{\hzrg}{HzRG}
\newcommand{\hzrgs}{HzRGs}
\newcommand{\spitzer}{{\em Spitzer}}
\newcommand{\chandra}{{\em Chandra}}
\newcommand{\kz}{\ensuremath{K-z}}
\newcommand{\chisq}{\ensuremath{\chi^2}}
\newcommand{\rchisq}{\ensuremath{\chi^2_\nu}}
\newcommand{\srca}{Source A}
\newcommand{\srcb}{Source B}
\newcommand{\srcc}{Source C}
\newcommand{\srcd}{Source D}
\newcommand{\srce}{Source E}
\newcommand{\srcaf}{J142631+341557}
\newcommand{\srcbf}{J142647+345851}
\newcommand{\srccf}{J142741+342324}
\newcommand{\srcdf}{J143036+334026}
\newcommand{\srcef}{J143258+342055}

\newcommand{\wphz}{W\,Hz\ensuremath{^{-1}}}
\newcommand{\raddata}{S.\ Croft et~al.\ (2008, in preparation)}

\shorttitle{High-$z$ protogalaxies}
\shortauthors{Croft \etal}

\begin{document}
\title{Radio-loud high-redshift protogalaxy candidates in Bo\"otes}
\author{Steve Croft\altaffilmark{1,2,3,4}, Wil van Breugel\altaffilmark{1,2}, Michael J.\ I.\ Brown\altaffilmark{5}, Wim de Vries\altaffilmark{1,3}, Arjun Dey\altaffilmark{6}, Peter Eisenhardt\altaffilmark{7}, Buell Jannuzi\altaffilmark{6}, Huub R\"ottgering\altaffilmark{8}, S.~A.~Stanford\altaffilmark{1,3}, Daniel Stern\altaffilmark{7} and S.~P.~Willner\altaffilmark{9}}
\altaffiltext{1}{Institute of Geophysics and Planetary Physics, Lawrence Livermore National Laboratory L-413, 7000 East Avenue, Livermore, CA 94550}
\altaffiltext{2}{University of California, Merced, P.O. Box 2039, Merced, CA 95344}
\altaffiltext{3}{Department of Physics, University of California at Davis, 1 Shields Avenue, Davis, CA 95616}
\altaffiltext{4}{University of California, Berkeley, 601 Campbell Hall \#3411, Berkeley, CA 94720}
\altaffiltext{5}{School of Physics, Monash University, Clayton, Victoria, Australia}
\altaffiltext{6}{National Optical Astronomy Observatory, 950 North Cherry Avenue, Tucson, AZ 85719}
\altaffiltext{7}{Jet Propulsion Laboratory, California Institute of Technology, Pasadena, CA 91109}
\altaffiltext{8}{Leiden Observatory, Leiden University, Niels Bohrweg 2, NL-2333 CA Leiden, The Netherlands}
\altaffiltext{9}{Harvard-Smithsonian Center for Astrophysics, 60 Garden Street, Cambridge, MA 02138}

%\tabletypesize{\scriptsize}

\begin{abstract}

We used the Near Infrared Camera on Keck I to obtain \ks-band images of four candidate high-redshift radio galaxies selected using optical and radio data in the NOAO Deep Wide-Field Survey in \bootes. Our targets have 1.4\,GHz radio flux densities greater than 1\,mJy, but are undetected in the optical to $\gtrsim 24$ Vega mag. % ($10^{-29}$\,\ecsh).
 Spectral energy distribution fitting suggests that three of these objects are at $z > 3$, with radio luminosities near the FR-I / FR-II break. The other has photometric redshift $z_{phot} = 1.2$, but may in fact be at higher redshift.

Two of the four objects exhibit diffuse morphologies in \ks-band, suggesting that they are still in the process of forming.
\end{abstract}


\section{Introduction}

High-redshift radio galaxies (\hzrgs) are rare objects in the cosmos, residing at the very brightest end of the radio luminosity function. Wide-area surveys have identified powerful radio galaxies out to very early cosmic epochs. However, since flux-limited surveys tend to probe increasingly luminous objects with increasing redshift, our knowledge of lower-luminosity \hzrgs\ at high redshift is sparse (TN\,J0924-2201 has a 1.4\,GHz flux density $S_{1400} = 73$\,mJy, despite being at $z = 5.19$; \citealt{wvb:99}). We have sought to remedy this situation with deep, multifrequency radio maps of the \bootes\ field (\citealt{devries:02}; \raddata; \S~\ref{sec:radio}). This field has substantial multiwavelength supporting data.

Many \hzrgs\ show diffuse rest-frame optical morphologies and other characteristics of ``protogalaxies'' \citep[\eg][]{pent:01}; they appear to form when Lyman-break-galaxy-sized clumps merge, and eventually evolve into large elliptical galaxies \citep{wvb:98}. Identifying lower-luminosity radio galaxies at early cosmic epochs can help our understanding of the nature of active galactic nucleus (AGN) hosts as a function of radio luminosity, and can help constrain cosmic evolution of the radio galaxy population.

By applying a cut in radio spectral index, $\alpha$ ($S_{\nu} \propto \nu^{\alpha}$), we can preferentially select higher-redshift objects, since a combination of intrinsic evolutionary effects and an observational ``$k$-correction'' tend to cause $\alpha$ to become steeper (\ie, more negative) with increasing redshift \citep[for details see, \eg][]{kc:91}. Indeed, 35\%\ of sources with $\alpha < -1.3$ and $S_{1400} > 10$\,mJy observed by \citet{db:uss} were found to have $z > 3$.

This paper presents \ks-band images of radio sources with steeper than normal radio spectra and no detections in deep optical data to $\gtrsim 24$\,mag (Vega) from the NOAO Deep Wide-Field Survey (NDWFS) \bootes\ field, in order to study the morphologies and properties of candidate \hzrgs\ that are fainter than ``typical'' \hzrgs.

\citet{higdon:05} performed a complementary study of optically-invisible radio sources in the \bootes\ field. They used the Very Large Array (VLA) at 1.4\,GHz to map a 0.5~\sqdeg\ area to a limiting sensitivity (at field center) of $\sim 15$\,\uJy\,beam$^{-1}$. Thirty six of their 377 radio sources (10\,\%) were not visible in the NDWFS data, although 90\%\ of these 36 had flux densities less than 1\,mJy; in contrast (\S~\ref{sec:ndwfs}) we consider sources brighter than 1\,mJy, and with relatively steep spectral indices, from our radio maps (\S~\ref{sec:radio}) covering an order of magnitude larger area. Like us, however, \citeauthor{higdon:05} conclude that most of their optically-invisible objects are AGNs at relatively high redshifts. The region mapped by \citeauthor{higdon:05} has very little spatial overlap with our 325\,MHz data, and none of their optically-invisible radio sources is present in our radio catalog.

Throughout this paper, we use Vega magnitudes, J2000 coordinates, and assume an \lcdmparm\ cosmology \citep{wmap}.

\section{Radio data} \label{sec:radio} 

Observations of a 4.9~\sqdeg\ region of the \bootes\ field were made with the VLA A-array at 325\,MHz during four runs from 2003 June--August. The CLEAN beam had a FWHM of 5\arcsec\ and the limiting sensitivity was $\sigma \sim 150$\,\ujy. The data analysis and reduction will be discussed in detail in an upcoming paper (\raddata). A catalog of radio sources (as opposed to components, \ie, a double-lobed radio galaxy counts as one source) was generated using the method described by \citet{devries:06}.

We matched the 325\,MHz catalog (\raddata) to the 1.4\,GHz Westerbork Synthesis Radio Telescope (WSRT) dataset of \citet{devries:02}, which has $\sigma \sim 28$\,\uJy, an ellipsoidal CLEAN beam of 13\arcsec$ \times 27$\arcsec, and an area of 6.68~\sqdegs.  The typical positional accuracy at 1.4\,GHz for sources in the high signal-to-noise ratio (S/N) regime (which is the case for our sources since their fluxes are $> 1$\,mJy) is $0\farcs44$ ($1\sigma$). Six hundred and fifty three sources (in the 4.9~\sqdeg\ where the two radio datasets overlap) were found to be in common and hence had measured radio spectral indices, $\alpha$. The properties of this matched catalog will be discussed by \raddata.

\section{Sample selection}
\label{sec:ndwfs}
To select \hzrg\ candidates, we combined deep radio data (\S~\ref{sec:radio}) with optical data from the NDWFS \citep{ndwfs}, a deep ($B_W = 25.3$, $R = 24.1$, $I = 23.6$, $K \sim 19$\,mag; $3\sigma$, 5\arcsec\ diameter -- \citealt{brodwin:06}) optical / near-IR survey in \bootes. The magnitude limits for point sources are deeper, but \hzrgs\ are likely to be resolved in the $\sim 1$\arcsec\ seeing of NDWFS, so we used 4\arcsec\ diameter optical apertures to determine whether or not a source is identified. We used the \citet{brown:07} $I$-band selected catalog, which has improved photometry for faint objects compared to the public NDWFS DR3 Source Extractor (SExtractor) catalog. In computing photometric redshifts, we measured photometry from the NDWFS optical data in 4\arcsec diameter apertures at each of the radio positions (\S~\ref{sec:photoz}). The NDWFS \bootes\ field is also covered by deep infrared, X-ray, and other observations (\S~\ref{sec:archival}), although these data were not available to us at the time of our Near Infrared Camera (NIRC) observations and were not used as part of the selection. 

\begin{figure}
\centering
\includegraphics[width=\linewidth,draft=false]{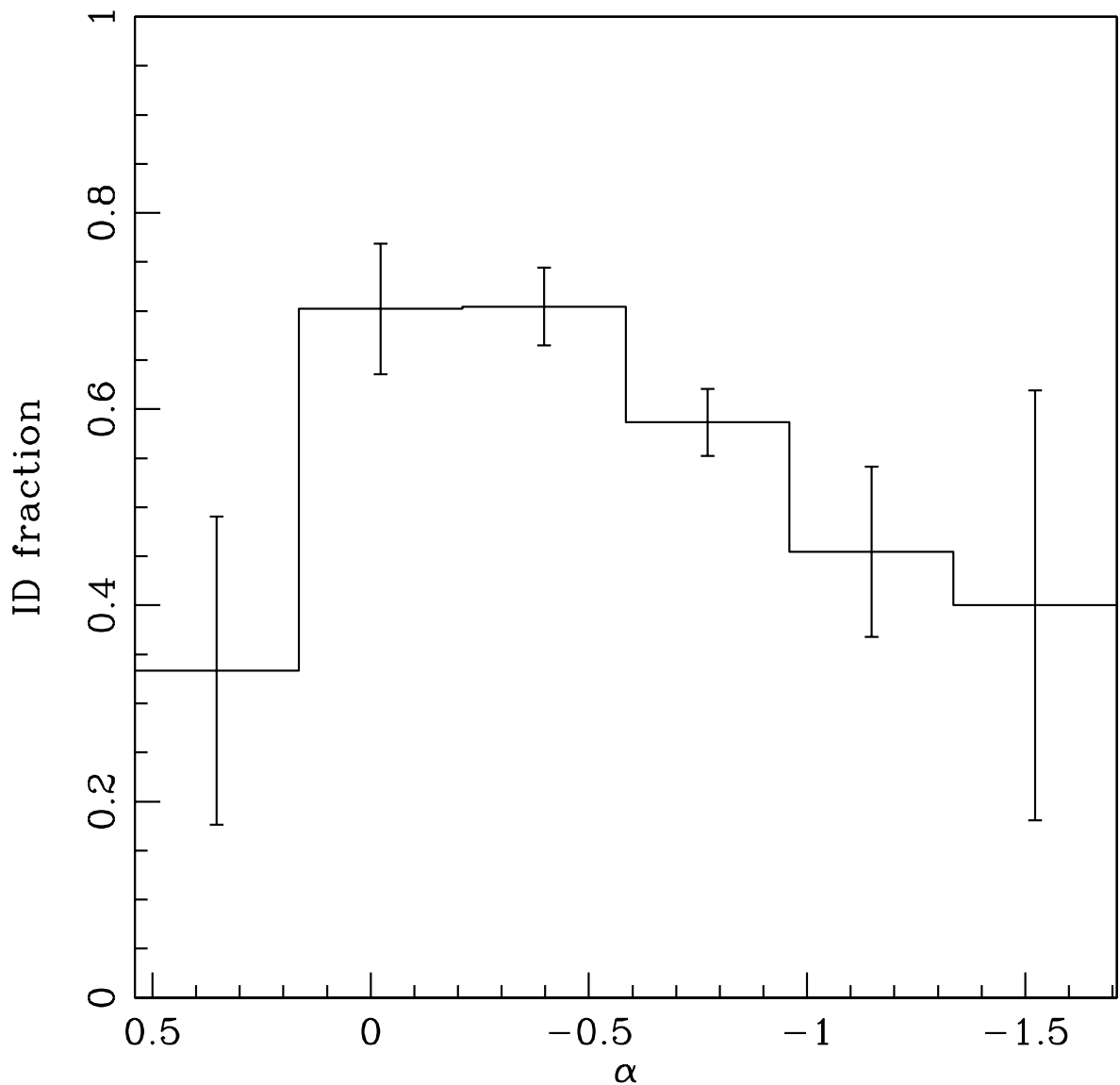}
\caption{\label{fig:spixid}Fraction of radio sources with $S_{1400} > 1$\,mJy that are identified in the NDWFS $R$-band data, as a function of radio spectral index, $\alpha$. Radio sources with inverted ($\alpha \gtrsim 0.2$) and steep spectral indices ($\alpha \lesssim -1$) have lower identification fractions than sources with ``normal'' spectral indices, suggesting that they are more distant and / or less luminous. This result will be discussed in more detail by \raddata.
}
\end{figure}

As noted by \citet{blumenthal:79}, \citet{tielens:79}, and others, radio sources with spectral indices that are steeper than normal (average $\alpha^{1400}_{325} \sim -0.7$) have a higher probability of being unidentified in optical surveys, a result that holds true for the combination of our radio data with the NDWFS $R$-band (Fig.~\ref{fig:spixid}; see \raddata\ for more details). We also see a similar drop in identification fraction for sources with inverted ($\alpha^{1400}_{325} \gtrsim 0.2$) spectra. At $S_{1400} > 1$\,mJy, the radio source counts are dominated by AGNs \citep{hopkins:00}. Pure starbursts with flux densities $>1$\,mJy must either be at relatively low redshift, or have an extremely high star-formation rate, and either ought to be easily seen in our deep optical data. One way to preferentially target \hzrgs, therefore, is to look for radio sources brighter than 1\,mJy that are optically faint or invisible, although of course even in these objects there may be some contribution to the radio luminosity from a starburst component. Deep $K$-band imaging can provide information on the morphologies of the \hzrgs, and the well-known \kz\ relation \citep[\eg,][]{ll:kz,db:kz,willott:kz} can be used to estimate their redshifts. 

\begin{figure*}
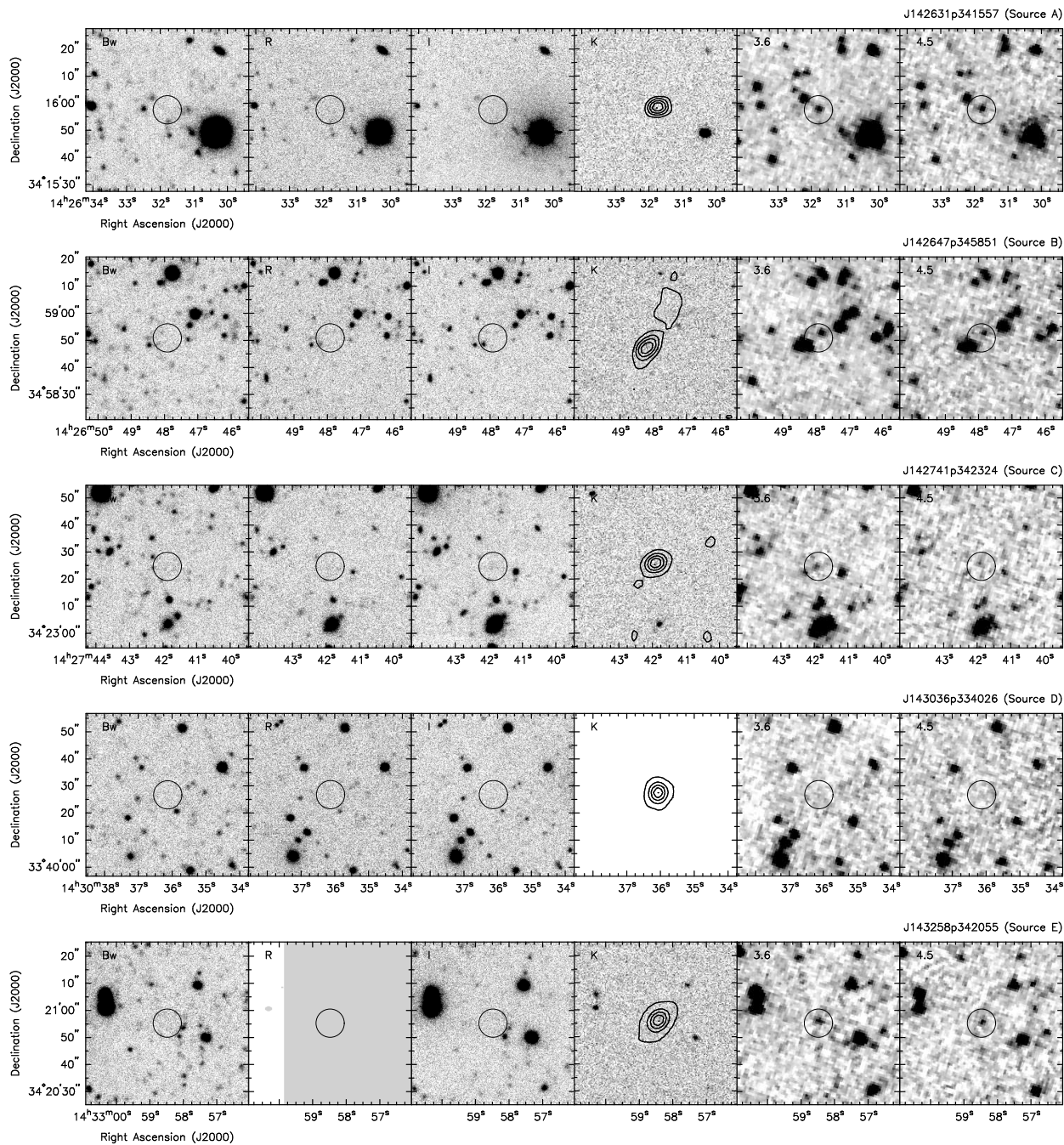

\centering
\includegraphics[width=\linewidth,draft=false]{f2a.eps}
\includegraphics[width=\linewidth,draft=false]{f2b.eps}
\includegraphics[width=\linewidth,draft=false]{f2c.eps}
\includegraphics[width=\linewidth,draft=false]{f2d.eps}
\includegraphics[width=\linewidth,draft=false]{f2e.eps}
\caption{\label{fig:postage}Postage stamps from NDWFS ($B_WRIK$) and the IRAC Shallow Survey (3.6 and 4.5\,\micron) for the five sources in our Keck imaging sample. A 5\arcsec\ radius circle is plotted at the 1.4\,GHz flux-weighted radio position \citep{devries:02} in each frame, except $K$, where the 325\,MHz radio map is overplotted. The lowest contour is at 0.4\,mJy\,beam$^{-1}$; the remaining contours are evenly spaced in flux density up to the peak flux density in each map.
}
\end{figure*}

Four hundred and thirty of the 434 radio sources with $S_{1400} > 1$\,mJy in the matched 325\,MHz / 1.4\,GHz sample are within the NDWFS imaging coverage. Seventy two of these 430 have no identifications in the NDWFS optical / IR data (which we define as no detections within 4\arcsec\ of the 1.4\,GHz radio position in $B_WRIK$ to 3$\sigma$ in a 4\arcsec\ diameter aperture). Some of these 72 objects have confusing or extended radio contours, are near bright stars, or appear to be associated with extended objects (\eg\ nearby galaxies larger than 8\arcsec\ in diameter where the radio position is offset from the optical centroid). We inspected finding charts of the NDWFS + 325\,MHz radio data, and selected a subsample of 14 sources that were unambiguously unidentified. These 14 sources have radio morphologies that appear unresolved at 325\,MHz, with no nearby extended radio emission, and are not near bright objects. Only one of the 14 has a steep spectrum, with $\alpha = -1.48$; four have $-0.98 \leq \alpha \leq -0.87$, seven have $- 0.80 \leq \alpha \leq -0.59$, and two have $\alpha > 0.1$. The five radio sources with the steepest spectra (Fig.~\ref{fig:postage}) were followed up with infrared imaging (\S~\ref{sec:irim}). Here we refer to these sources as A -- E; see Table~\ref{tab:obs} for positions and names.

\begin{deluxetable}{llllllllll}
\rotate
\tablewidth{0pt}
\tabletypesize{\scriptsize}
\tablecaption{\label{tab:obs} Observation log}
\tablehead {
\colhead{Source name} & \colhead{J2000 ID\tablenotemark{a}} & \colhead{1.4\,GHz RA} & \colhead{1.4\,GHz Dec} & \colhead{\ks\ RA} & \colhead{\ks\ Dec} & \colhead{4.5\,\micron\ RA} & \colhead{4.5\,\micron\ Dec} & \colhead{NIRC exp (s)} & \colhead{LRIS exp (s)}
}
\startdata
\srca & \srcaf & \hms{14}{26}{31.75} & \dms{34}{15}{57.5} & \hms{14}{26}{31.8} & \dms{34}{15}{57} & \hms{14}{26}{31.75} & \dms{34}{15}{58.5} & 3840 & 0\\ 
\srcb & \srcbf &  \hms{14}{26}{47.87} & \dms{34}{58}{51.0} & \hms{14}{26}{47.8} & \dms{34}{58}{53} & \hms{14}{26}{47.78} & \dms{34}{58}{53.3} & 1920 & 0\\
\srcc & \srccf & \hms{14}{27}{41.84} & \dms{34}{23}{24.7} & \hms{14}{27}{41.9} & \dms{34}{23}{25} & \hms{14}{27}{41.85} & \dms{34}{23}{26.3} & 3840 & 3600 \\ 
\srcd & \srcdf & \hms{14}{30}{36.09} & \dms{33}{40}{26.6} & \multicolumn{4}{c}{No ID - radio source is a lobe} & 1920 & 0\\ 
\srce\tablenotemark{b} & \srcef & \hms{14}{32}{58.44} & \dms{34}{20}{55.4} & \hms{14}{32}{58.5} & \dms{34}{20}{56} & \hms{14}{32}{58.44} & \dms{34}{20}{56.0} & 1920 & 4680\\
\enddata
\tablenotetext{a}{J2000 identifiers are from the 1.4\,GHz catalog of \citet{devries:02}. All five sources listed correspond to 325\,MHz counterparts without ambiguity. The \citeauthor{devries:02} catalog is denoted [DMR2002] by SIMBAD.} 
\tablenotetext{b}{Detected in \xbootes\ X-ray data (\S~\ref{sec:xray})}
\end{deluxetable}

Despite the efficiency of the method used for combining multiple radio source components in order to eliminate radio lobes from the catalog (\S~\ref{sec:radio}), inevitably a few lobes may slip through. One such object turns out to be \srcd; it was undetected in our NIRC observations, and later turned out to be undetected in any of the IRAC bands. When we overplotted our VLA data, as well as data from the NRAO VLA Sky Survey (NVSS), on a wide-field Digitized Sky Survey (DSS) image, we found that this source is in fact a lobe of a large (4\farcm5\ diameter) FR-II  that has a faint $R$-band identification in the DSS. Additionally, after the observations were taken, we replotted the radio data for \srcb\ on the NDWFS images, including a lower contour than in our original plots, and noticed that the radio emission appeared extended (Fig.~\ref{fig:postage}), and so this object ought not to have been included in our original sample of 14 objects with unresolved radio emission and no NDWFS matches. Nevertheless, the radio emission appears to be associated with an IR-bright source which meets our criteria for optical faintness. \srcd\ (the large FR-II) will not be discussed further here; the remaining four fields resulted in the detection of candidate \hzrgs, as discussed in \S~\ref{sec:discuss}.

\section{Keck NIRC / LRIS Observations}
\label{sec:irim}

\label{sec:nirc}Imaging observations were made in \ks-band, using the NIRC \citep{nirc}, --- a $256 \times 256$ InSb array with 0\farcs15 pix$^{-1}$ --- on Keck I during two half nights on UT 2004 April 29 and 30. A 16-position ($4 \times 4$) dither, with dither spacing 3\arcsec\ and 60\,s per exposure ($6 \times 10\,s$) was used. Three fields were observed for four dithers (3840\,s) and the remaining two fields observed for two dithers (1920\,s), as summarized in Table~\ref{tab:obs}. Seeing was 0\farcs5 -- 0\farcs6 and the sky was clear. The data were reduced in the standard manner using DIMSUM\footnote{\url{http://iraf.noao.edu/iraf/ftp/contrib/dimsumV2}} in IRAF. Astrometry was bootstrapped from the NDWFS data; the small offsets between the \ks\ and 4.5\,\micron\ positions (Table~\ref{tab:obs}) suggest that our NIRC astrometry is accurate to $\sim 1$\arcsec. Photometry was performed in a 3\farcs5 diameter aperture; the \ks\ magnitudes are presented in Table~\ref{tab:targets}, where we also present the measured NDWFS and ISS photometry (\S~\ref{sec:photoz}).

\begin{deluxetable}{llllllllllllll}
\rotate
\tablewidth{0pt}
\tabletypesize{\scriptsize}
\tablecaption{\label{tab:targets} Properties of the \hzrg\ candidates}
\tablehead {
\colhead{ID} & \colhead{$\alpha^{1400}_{325}$} & \colhead{$S_{1400}$ (mJy)} & \colhead{$B_W$ \tablenotemark{a}} & \colhead{$R$ \tablenotemark{a}} & \colhead{$I$ \tablenotemark{a}} & \colhead{$z$ \tablenotemark{b}} & \colhead{$J$ \tablenotemark{c}} & \colhead{$K_s$ \tablenotemark{d}} & \colhead{ [3.6] \tablenotemark{e}} & \colhead{ [4.5] \tablenotemark{e}} & \colhead{ [5.8] \tablenotemark{e}} & \colhead{ [8.0] \tablenotemark{e}}
}
\startdata
\srca & -1.48 & $1.08 \pm 0.04$ & $27.60 \pm 1.51$ & $25.06 \pm 0.64$ & $22.50 \pm 0.29$  & $> 23.1$ & $> 20.8$   & $20.94 \pm 0.10$ & $18.12 \pm 0.14$ & $17.34 \pm 0.14$ & $15.84 \pm 0.31$  & $15.98 \pm 0.57$ \\
\srcb & -0.89 & $8.93 \pm 0.04$ & $28.25 \pm 1.76$ & $25.91 \pm 0.91$ & $27.87 \pm 3.36$  & \nodata & \nodata & $19.63 \pm 0.05$ & $18.03 \pm 0.13$ & $17.58 \pm 0.18$ & $19.10 \pm 6.05$ & $15.75 \pm 0.47$ \\
\srcc & -0.98 & $3.61 \pm 0.03$ & $26.70 \pm 0.73$ & $25.32 \pm 1.12$ & $24.40 \pm 0.81$  & $> 22.9$ & \nodata  & $21.38 \pm 0.13$ & $19.23 \pm 0.37$ & $18.51 \pm 0.40$ & $17.27 \pm 1.13$ & $16.02 \pm 0.60$ \\
\srce & -0.87 & $16.27 \pm 0.03$ & $26.37 \pm 0.76$ & $25.73 \pm 1.24$ & $24.48 \pm 1.10$ & $> 23.2$ & $> 20.6$    & $20.99 \pm 0.13$ & $18.45 \pm 0.19$ & $17.88 \pm 0.23$ & $17.08 \pm 0.95$ & $16.42 \pm 0.85$ \\
\enddata
\tablecomments{We report measured photometric uncertainties, even when these are large (\eg 6.05\,mag for \srcb\ at 5.8\,\micron), since these were the values used in photometric redshift fitting (\S~\ref{sec:photoz}). All magnitudes are on the Vega system.}
\tablenotetext{a}{NDWFS (\S~\ref{sec:ndwfs})}
\tablenotetext{b}{z\bootes\ (\S~\ref{sec:zbootes})}
\tablenotetext{c}{FLAMEX (\S~\ref{sec:flamex})}
\tablenotetext{d}{NIRC (\S~\ref{sec:nirc})}
\tablenotetext{e}{IRAC Shallow Survey (\S~\ref{sec:iss})}
\end{deluxetable}

Two of the four \hzrg\ candidates were observed spectroscopically. The slitmasks were filled with secondary targets selected by their IR colors from the AGN ``wedge'' of \citet{stern:05}, along with a few HzRG candidates; these will be discussed in future papers. 
These observations used the Low Resolution Imaging Spectrometer \citep[LRIS;][]{lris} on Keck I, with the D560 dichroic and a slit width of 1\farcs3. On the blue side, a 400 line mm$^{-1}$ grism, blazed at 3400\,\AA\ was employed, giving 1.09 \AA\ / pixel and spectral resolution 8.1\,\AA. On the red side, a 400 line mm$^{-1}$ grating, blazed at 8500\,\AA\ was used, giving 1.86 \AA\ / pixel and spectral resolution 7.3\,\AA. This setup gives spectral coverage of $\sim 3150-9400$\,\AA, with some variation from slitlet to slitlet. \srcc\ was observed on UT 2005 June 4 for 3600\,s ($2 \times 1800$\,s) under clear skies, in 1\farcs4 seeing. \srce\ was observed on UT 2005 June 6 for 4680\,s ($3 \times 1560$\,s) in 0\farcs8 seeing and with clear skies. The observations are summarized in Table~\ref{tab:obs}.

The data were reduced in the standard manner using BOGUS\footnote{\url{https://zwolfkinder.jpl.nasa.gov/$\sim$stern/homepage/bogus.html}} in IRAF, and spectra extracted in a 1\farcs5 wide aperture for many of the secondary targets. The primary targets were undetected in both the red and blue side two-dimensional spectra, and for these no extractions were performed.

\section{Other data}
\label{sec:archival}

\subsection{X-ray}
\label{sec:xray}

The \bootes\ field is covered by the \chandra\ X-ray observations of \citet{xbootesmap}. The resulting catalogs, published by \citet{xbootes} and \citet{xbootesopt}, reach a limiting full-band ($0.5 - 7.0$\,keV) flux of $7.8 \times 10^{-15}$\,\esc\ for on-axis sources with four counts. We checked for counterparts to our targets, and found that \srce\ was detected\footnote{(R.A., decl.) = (\hmst{14}{32}{58.49}, \dmst{34}{20}{54.77}) $\pm 1.25$\arcsec} (five counts in the full band, corresponding to $1.13 \pm 0.71 \times 10^{-15}$\,\esc ). This source is very likely to be a powerful high redshift and / or optically obscured quasar simply on the basis of its large X-ray-to-optical flux ratio \citep{xbootesopt}. 

Intriguingly, an object in the field of \srcb\ is also detected in X-rays with 5 counts in the full band ($1.17 \pm 0.71 \times 10^{-15}$\,\esc ). The X-ray position\footnote{(R.A., decl.) = (\hmst{14}{26}{48.41}, \dmst{34}{58}{47.37}) $\pm 1.60$\arcsec} is close to the peak of the 325\,MHz radio contours, but offset (by 7\farcs0) from the flux-weighted 1.4\,GHz position. However, it appears that the X-ray and radio sources are two distinct objects; if they were the same (or if the radio source were associated with the object nearest the peak of the radio contours), the radio source would be extremely core dominated, which is not supported by the comparatively steep spectral index. Rather, it appears that the radio source is an extended, double-lobed source associated with the $I$-band dropout at the position given in Table~\ref{tab:obs}, and that the X-ray source is associated with the optical object\footnote{(R.A., decl.) = (\hmst{14}{26}{48.45}, \dmst{34}{58}{48.47})} nearest (1\farcs2) the X-ray position, and clearly visible in the NDWFS $I$-band (Fig.~\ref{fig:postage}) and NIRC \ks-band (Fig.~\ref{fig:nirccon}) data.

Both of the X-ray sources have too few counts to reliably determine a hardness ratio (three counts in the soft band and two in the hard band in both cases).

The other \hzrg\ candidates are not detected in \xbootes, although they lie within the coverage area. Since the catalogs of \citet{xbootes} and \citet{xbootesopt} contain only sources with four or more counts, it is possible that they may be X-ray sources, but be insufficiently bright for inclusion in the full catalog. None of our sources is within the coverage area of the more sensitive ($\sim 10^{-16}$\,\esc ), single-pointing \chandra\ X-ray observations of \citet{lala}. 

\subsection{Infrared}

\label{sec:zbootes} The z\bootes\ survey \citep{zbootes} provides $z$-band data for part of the \bootes\ field. We checked for counterparts to our four \hzrg\ candidates, and found that three were undetected at the z\bootes\ flux limits (Table~\ref{tab:targets}), while \srcb\ was outside the z\bootes\ coverage area.

\begin{figure}
\centering
\includegraphics[width=\linewidth,draft=false]{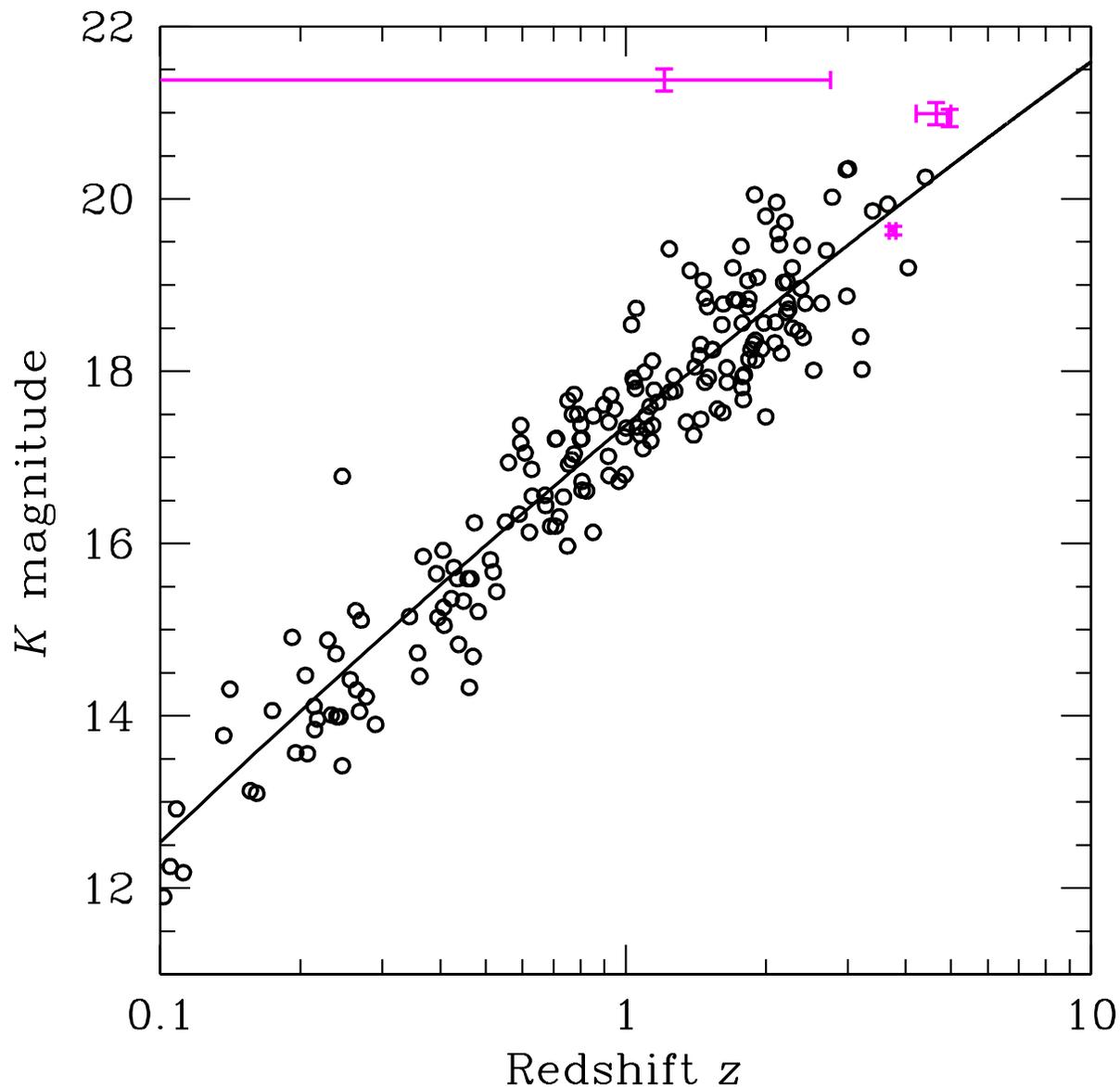}
\caption{\label{fig:kz}\kz\ diagram of \citet{willott:kz} (circles represent radio galaxies with spectroscopic redshifts from the 3CRR, 6CE, 6C$^{\star}$, and 7CRS surveys). Our \hzrg\ candidates are plotted as points with error bars, at the redshifts obtained from photometric redshift fitting. Vertical error bars represent the measured uncertainties in the \ks\ band magnitudes (where we assume $K - \ks$ is negligible), and horizontal error bars represent the 68\%\ confidence interval in redshift from HyperZ as tabulated in Table~\ref{tab:pz}.
}
\end{figure}

\label{sec:flamex}FLAMEX \citep{flamex} provides $J$- and \ks-band catalogs for part of the \bootes\ field. The \ks-band catalogs are deeper than the $K$-band catalogs used for our selection, but our measured NIRC \ks-band fluxes (for the three sources in regions covered by FLAMEX) are below the FLAMEX \ks-band limit ($\ks \sim 19.3$\,mag), so these sources are too faint to appear in the \ks-band FLAMEX catalog. The $J$-band FLAMEX limit ($J \sim 21$\,mag) is insufficient to put anything but the weakest constraints on $J - \ks$ colors, and in any case, two of our sources are outside the FLAMEX coverage area. In future, refining our selection criteria by requiring non-detections in FLAMEX will further increase our efficiency at finding \hzrgs, since $K > 19$ corresponds to $z \gtrsim 2$ for sources on the \kz\ relation (Fig.~\ref{fig:kz}; \S~\ref{sec:discuss}).

\label{sec:iss}We checked the IRAC \citep{irac} catalogs of the IRAC Shallow Survey \citep{iss} for infrared counterparts to our \hzrg\ candidates. All were detected at greater than $2\sigma$ significance at 3.6 and 4.5\,\micron, and one also at 5.8\,\micron. We report the catalog 5\arcsec\ aperture magnitudes in Table~\ref{tab:targets}, and use these measurements to assist with spectral energy distribution (SED) fitting (\S~\ref{sec:discuss}). Table~\ref{tab:targets} shows measured values, even in cases where the photometric uncertainties are large, since real measurements from the data reflect the relative weights given to the data points in the \chisq\ photometric redshift fitting. All objects are pointlike at IRAC resolution ($\sim 2\arcsec$).

\label{sec:mips}We downloaded the post-BCD 24\,\micron\ mosaics from the \spitzer\ archival observations of Soifer~et~al.\ for the fields surrounding the four \hzrg\ candidates, and performed quick-look photometry using the published zero points. Three of our objects were undetected (to a flux limit of $\sim 0.2$\,mJy) but \srca\ (the ultra-steep spectrum source) was detected with $S_{24\,\micron} = 0.43 \pm 0.13$\,mJy. 

\section{Discussion} \label{sec:discuss}

\begin{figure}
\centering
\includegraphics[width=\linewidth]{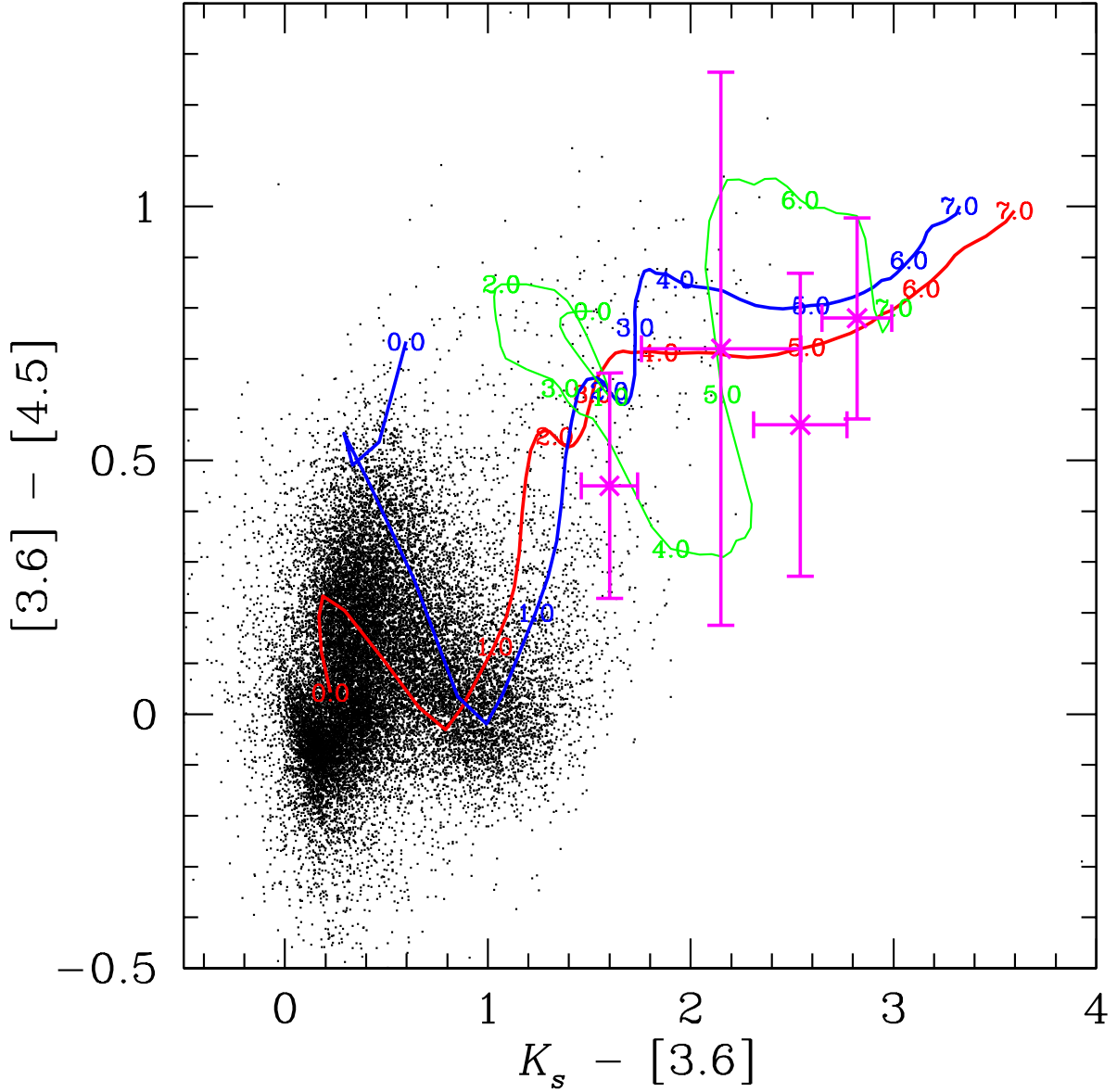}
\caption{\label{fig:cc} 
[3.6] $-$ [4.5] versus \ks\ $-$ [3.6] diagram for our four \hzrg\ candidates (magenta crosses with photometric error bars), compared to the colors of field sources in \bootes\ with at least a $5\sigma$ detection at NDWFS $K$-band (where we assume that $K - \ks$ is negligible), 3.6\,\micron\ and 4.5\,\micron\ (black points). Also plotted are the colors of non-evolving galaxy templates from \citet{polletta:07}, redshifted from $z = 0$ to $z = 7$ in steps of $z = 0.1$ (with redshifts marked in steps of $z = 1.0$). The red line is a 2\,Gyr old elliptical template, the blue line is Arp\,220, and the green line is a QSO2.
}
\end{figure}

We have \ks\ and IRAC detections and optical upper limits for our \hzrg\ candidates (Table~\ref{tab:targets}), which we use to make somewhat crude redshift estimates. \ks\ $-$ [3.6], which straddles the Balmer and / or 4000\,\AA\ breaks for the two $z_{phot} > 4$ sources from Table~\ref{tab:pz}, provides quite tight redshift constraints. The position of our sources on a diagram of [3.6] $-$ [4.5] versus \ks\ $-$ [3.6] color (Fig.~\ref{fig:cc}) suggests that three are at $3 \lesssim z \lesssim 5$.

\begin{deluxetable}{llllllll}
\rotate
\tablewidth{0pt}
\tabletypesize{\scriptsize}
\tablecaption{\label{tab:pz} Properties of the \hzrg\ candidates from photometric redshift fitting}
\tablehead {
\colhead{ID} & \colhead{$z_{phot}$} & \colhead{$z$ (68\% confidence)} & \colhead{$z$ (99\% confidence)} & \colhead{Template} & \colhead{$A_V$} & \colhead{M($B_{rest}$)} & \colhead{Rest-frame 1.4\,GHz $L$ ($10^{26}$\,\wphz)}
}
\startdata
%\srca & 4.97 & 4.92 - 4.98 & 3.94 - 5.69 & Arp220  & 0.00 & -24.65 & 6.80 \\ 
%\srcb & 3.76 & 3.68 - 3.81 & 1.77 - 4.08 & QSO2    & 0.30 & -24.30 & 10.4 \\
%\srcc & 1.21 & 0.00 - 2.76 & 0.00 - 6.99 & Mrk231  & 0.00 & -19.07 & 0.301 \\ 
%\srce & 4.65 & 4.21 - 4.99 & 0.00 - 6.99 & Arp220  & 0.00 & -24.05 & 29.7 \\ 
\srca & 4.97 & 4.92 - 4.98 & 3.94 - 5.69 & Arp220  & 0.00 & -24.67 & 6.94 \\ 
\srcb & 3.76 & 3.68 - 3.81 & 1.77 - 4.08 & QSO2    & 0.30 & -24.32 & 10.5 \\
\srcc & 1.21 & 0.00 - 2.76 & 0.00 - 6.99 & Mrk231  & 0.00 & -19.05 & 0.298 \\ 
\srce & 4.65 & 4.21 - 4.99 & 0.00 - 6.99 & Arp220  & 0.00 & -24.07 & 30.2 \\ 
\enddata
\tablecomments{Precise values of $z_{phot}$ are reported, even when the associated uncertainties are large, so that the reader may know the exact redshifts used to compute rest-frame properties of the sources}
\end{deluxetable}

\label{sec:photoz}We also obtained photometric redshifts using the publicly available software HyperZ \citep[Version 1.2,][]{hyperz}. In order to obtain more stringent limits on the fluxes of our objects than the 5\arcsec\ aperture limits used in the selection criteria, we remeasured photometry from the NDWFS $B_WRI$ images, in a 4\arcsec\ aperture at the position of the radio centroid. Empirical photometric errors were derived using the method of \citet{brown:07}, a Monte Carlo technique which places apertures on random regions of sky. This provides more accurate estimates of uncertainties (verified by the addition of simulated galaxies to the real data) than methods which assume uncorrelated Poisson background noise and no source confusion. Although some of the resulting measurements have large uncertainties (Table~\ref{tab:targets}), they help to constrain the resulting photometric redshifts better than using the published survey limiting magnitudes as upper limits. In cases where they are large, the quoted uncertainties provide a real measure of the variations in the background values due to neighboring objects. Since the uncertainties are used essentially as weights in the \chisq\ photometric fitting, large uncertainties downweight the effect of bands where the object flux is poorly constrained, as opposed to assigning these bands too high a weight by setting the fluxes to zero and obtaining the uncertainty from the survey flux limit. However, even when we re-run the photometric redshift fitting using the survey limits for those sources with uncertainties larger than 1\,mag, we find that there is essentially no change in the best-fit redshift (the largest change is for \srce, whose best-fitting template changes from Arp220 to a 2\,Gyr old elliptical, and whose redshift changes by just 0.04), although the confidence intervals become marginally less well constrained. For the $z$- and $J$-bands, where we only have access to catalogs and not to the original images (in order to remeasure photometry at the radio positions), we set the fluxes to zero and derive the associated uncertainties from the published survey limiting magnitudes. We find that whether or not the $z$- and $J$-band data points are included has essentially no effect on the best-fitting redshifts and confidence intervals (due to the large uncertainties in these bands). 

\begin{figure*}[tb]
\centering
\includegraphics[height=0.4\linewidth]{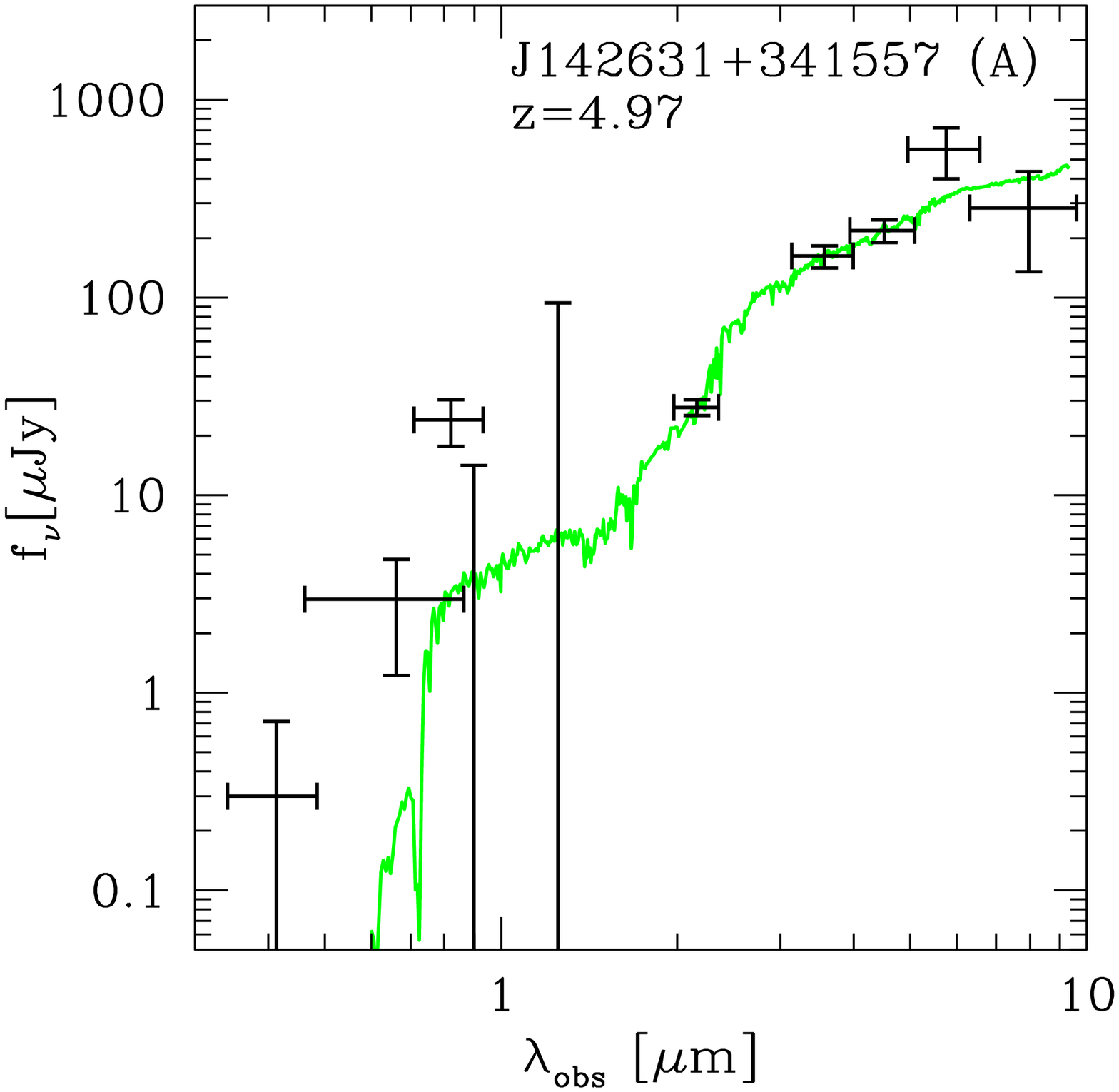}\hspace{0.03\linewidth}%
\includegraphics[height=0.4\linewidth]{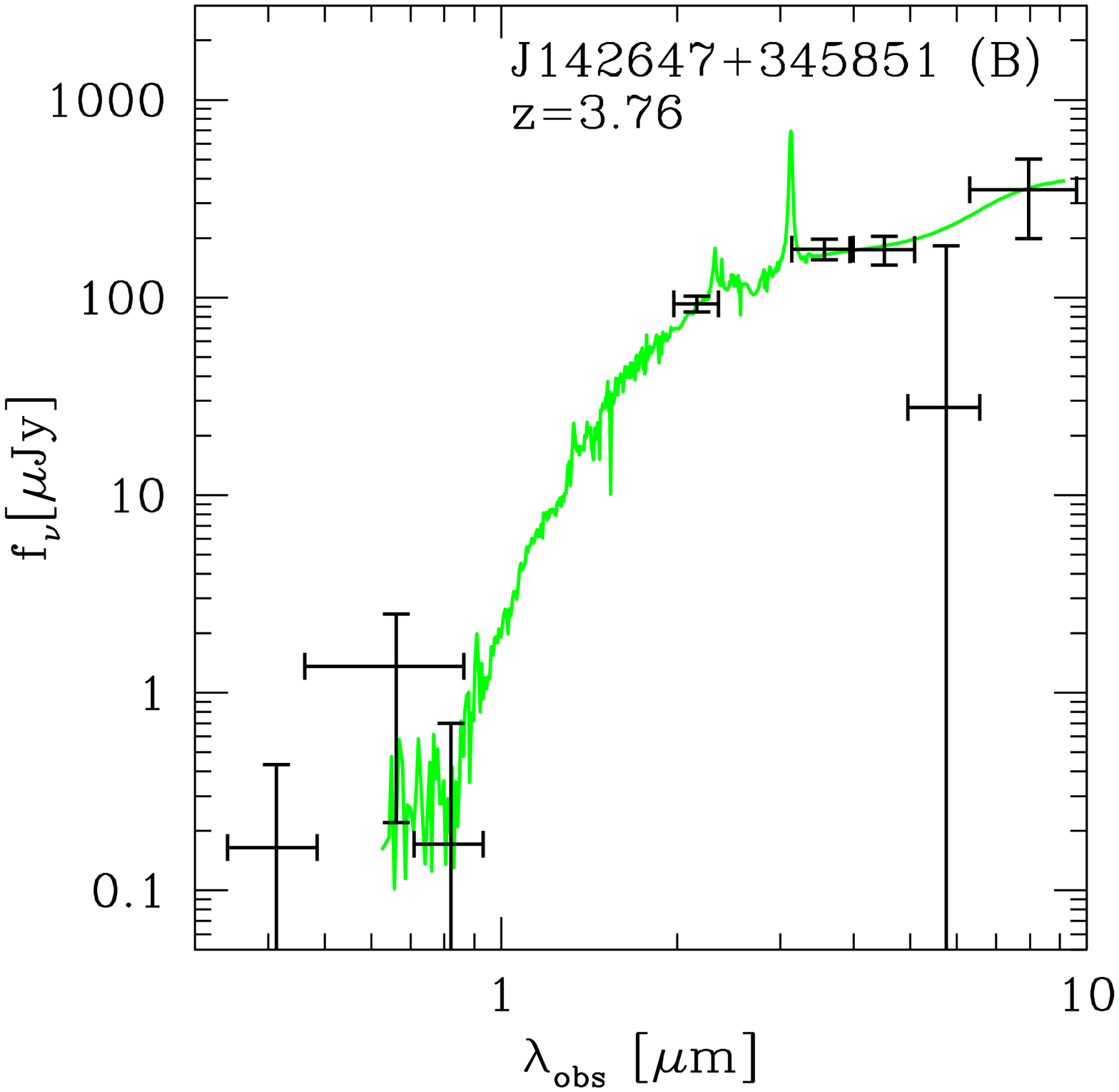}\hspace{0.03\linewidth}
\includegraphics[height=0.4\linewidth]{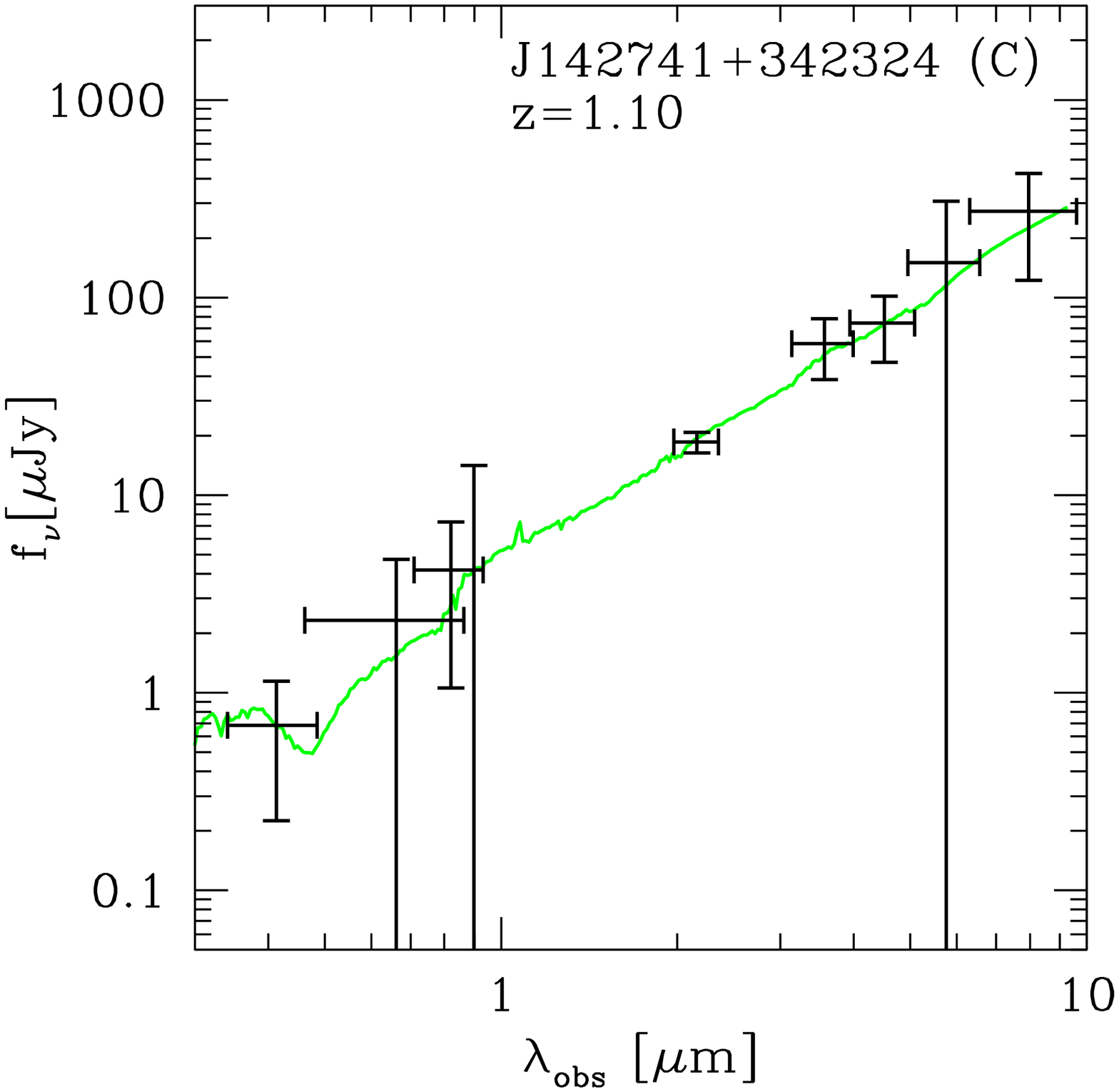}\hspace{0.03\linewidth}%
\includegraphics[height=0.4\linewidth]{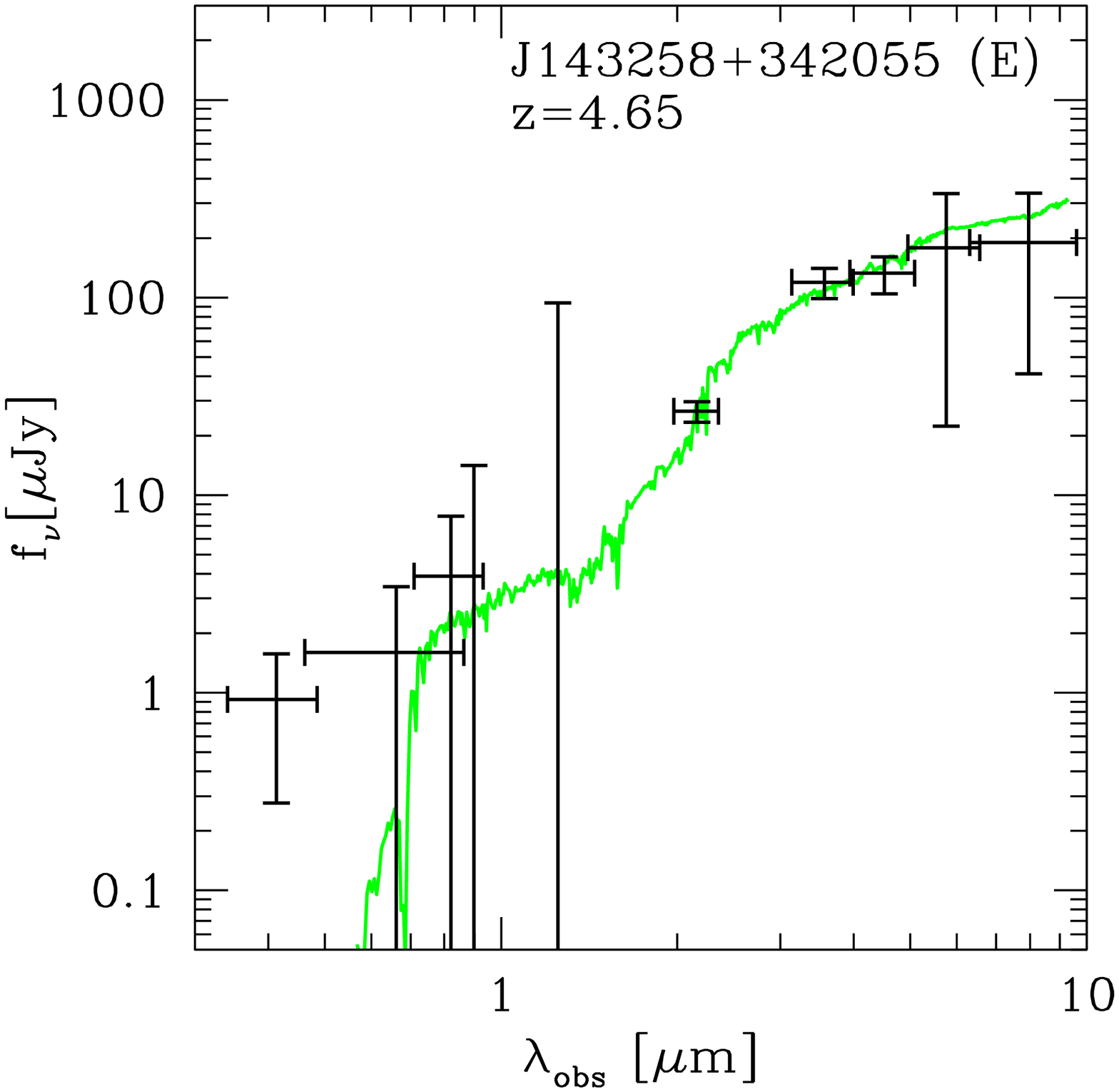}\hspace{0.03\linewidth}
\caption{\label{fig:sed}SEDs of the \hzrg\ candidates. The best-fitting templates are shown plotted at the redshift which gave the lowest \rchisq\ in SED fitting. We also note the best-fit redshift in each panel; note, however, that the redshifts are not very tightly constrained (Table~\ref{tab:pz}).
}
\end{figure*}

The $I$-band point for \srca\ seems rather high (Fig.~\ref{fig:sed}); inspection of the NDWFS $I$-band postage stamp in Fig.~\ref{fig:postage} suggests that this is probably due to the unrelated saturated star to the southwest. In fact, even though \srca\ fulfils our initial selection criterion (no source in the \citealt{brown:07} catalog closer than 4\arcsec), the aperture magnitude measured at the radio position (Table~\ref{tab:targets}) is brighter than the $3\sigma$ survey flux limits quoted in \S~\ref{sec:ndwfs}, strongly suggesting that light from a nearby object is contributing to the measured aperture magnitude (the NDWFS $I$-band point-spread function (PSF) has wings which can cause problems for aperture photometry at surprisingly large radii). If the photometric redshift is correct, there could also be some contribution to the flux in this band from \lya\ at $z \sim 5$, although an $I = 22.5$ source ought to be visible in the postage stamps (Fig.~\ref{fig:postage}). Excluding the $I$-band point from the SED fitting for this object has no effect on the best-fitting redshift, and little effect on the confidence intervals.

Instead of using the templates provided with HyperZ, we used a selection of templates (the QSO2, Arp220, Mrk231, Ell2, Ell5 and Ell13 templates; the latter three are 2, 5, and 13\,Gyr old ellipticals) from \citet{polletta:07}, which do a better job of reproducing IR fluxes. The effects of dust extinction were simulated using the prescriptions of \citet{calzetti:00}, with extinction $A_V$ allowed to vary between $0.0$ and $2.0$ mag (although some of the templates already incorporate the effects of some intrinsic dust extinction). $B$-band absolute magnitude, $M_B$ was constrained to be $-28.8 \leq M_B \leq -19$. Because of the faintness of these objects in the optical, and the corresponding large uncertainties for the optical data points compared to the IR, the range of plausible redshifts is rather large (Table~\ref{tab:pz}). However, we can still estimate the best-fitting redshifts, $z_{phot}$. The template SEDs for the best-fitting models are shown in Fig.~\ref{fig:sed}.

Three of the four \hzrg\ candidates have $z_{phot} > 3$, and the fourth is rather poorly constrained (due to low S/N) to $z_{phot} = 1.2$ (Table~\ref{tab:pz}). We report precise values for $z_{phot}$ in order that the reader may know what values were used for the calculation of the rest-frame radio luminosities and other such properties. However, as can be seen from the confidence limits in redshift reported in Table~\ref{tab:pz}, we cannot constrain redshifts for these objects very accurately using SED fitting, given the large number of bands with non-detections or large photometric errors. \srcb\ was not observed spectroscopically, so we cannot say whether or not the emission lines seen in the template (Fig.~\ref{fig:sed}) are really present in its spectrum. In the event of galaxy-wide dust extinction these spectral lines would be less prominent, but this ought not to affect our determination of $z_{phot}$ too much.

\begin{figure}
\includegraphics[width=\linewidth]{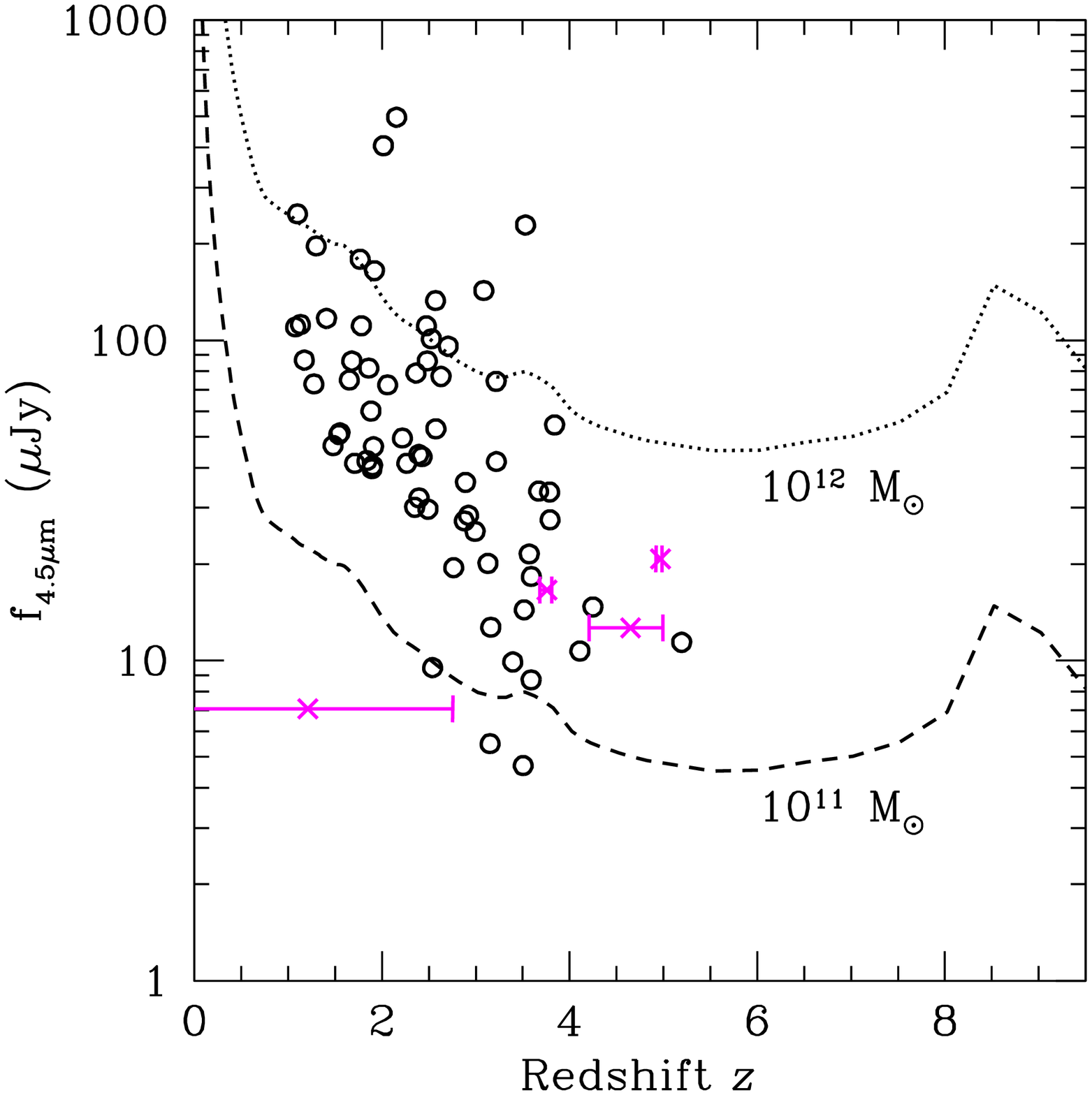}
\caption{\label{fig:4p5z}
4.5\,\micron\ flux density versus redshift for \hzrgs\ with spectroscopic redshifts from \citet{seymour:07} (circles), and our \hzrg\ candidates (crosses, plotted at the best-fit photometric redshift with 68\%\ confidence intervals shown). The dashed and dotted lines are evolutionary tracks for elliptical galaxies with formation redshift $z_f = 10$ and masses $10^{11}$ and $10^{12}$\,\msun, respectively.
}
\end{figure}

We can also use the 4.5\,\micron\ flux densities alone as an alternative to the \kz\ relation. \citet{seymour:07} studied a sample of 69 \hzrgs\ with spectroscopic redshift $1 < z < 5.2$ and a wide range of radio luminosities. They showed that the rest-frame $H$-band light for these objects is well described by models of elliptical galaxies with formation redshift $z_f = 10$ and masses between $10^{11}$ and $10^{12}$\,\msun. In Fig.~\ref{fig:4p5z}, we plot the measured values of the 4.5\,\micron\ flux density from \citeauthor{seymour:07} against redshift, along with tracks representing \citet{bc:93} models with $z_f = 10$ and solar metallicity, normalized to $10^{11}$ and $10^{12}$\,\msun. Plotting our \hzrg\ candidates in this figure at their best-fit photometric redshifts, these redshifts seem reasonable, although the lower bounds on the redshift (assuming our objects have masses $>10^{11}$\,\msun) are still quite weak ($z \gtrsim 2$). \srcc, with $z_{phot} = 1.21$, falls below the $10^{11}$\,\msun\ line, suggesting that it is either unusually low mass, or that the SED fitting underestimates the true redshift (certainly possible given the wide confidence interval and low S/N for this source). \srcc\ is the only outlier on the \kz\ plot (Fig.~\ref{fig:kz}), again suggesting it is either very underluminous, or is in fact at higher redshift than suggested by the SED fits.

\begin{figure*}[p]
\centering
\includegraphics[width=0.45\linewidth,draft=false]{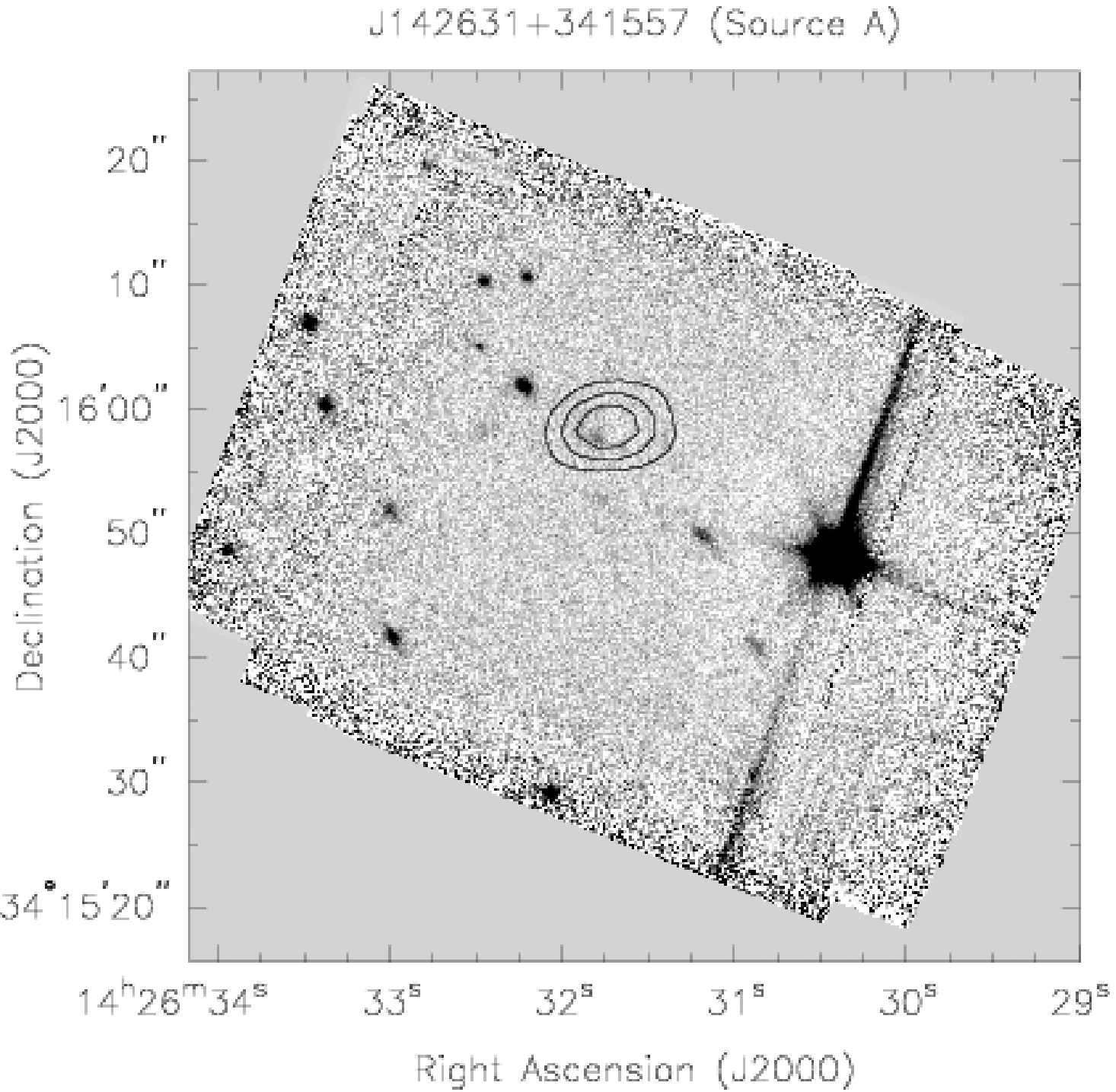}\hspace{0.03\linewidth}%
\includegraphics[width=0.45\linewidth,draft=false]{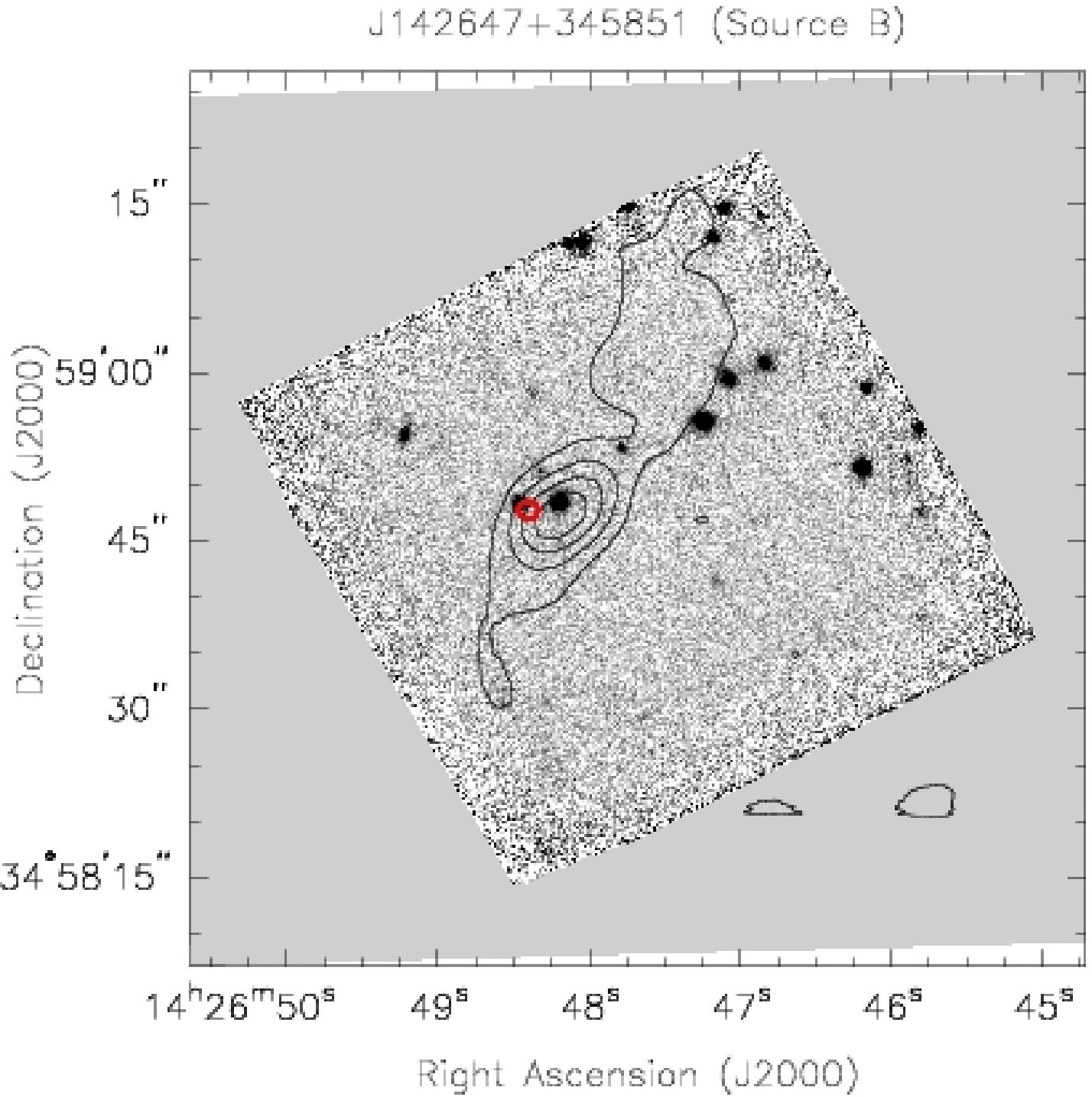}\hspace{0.03\linewidth}
\includegraphics[width=0.45\linewidth,draft=false]{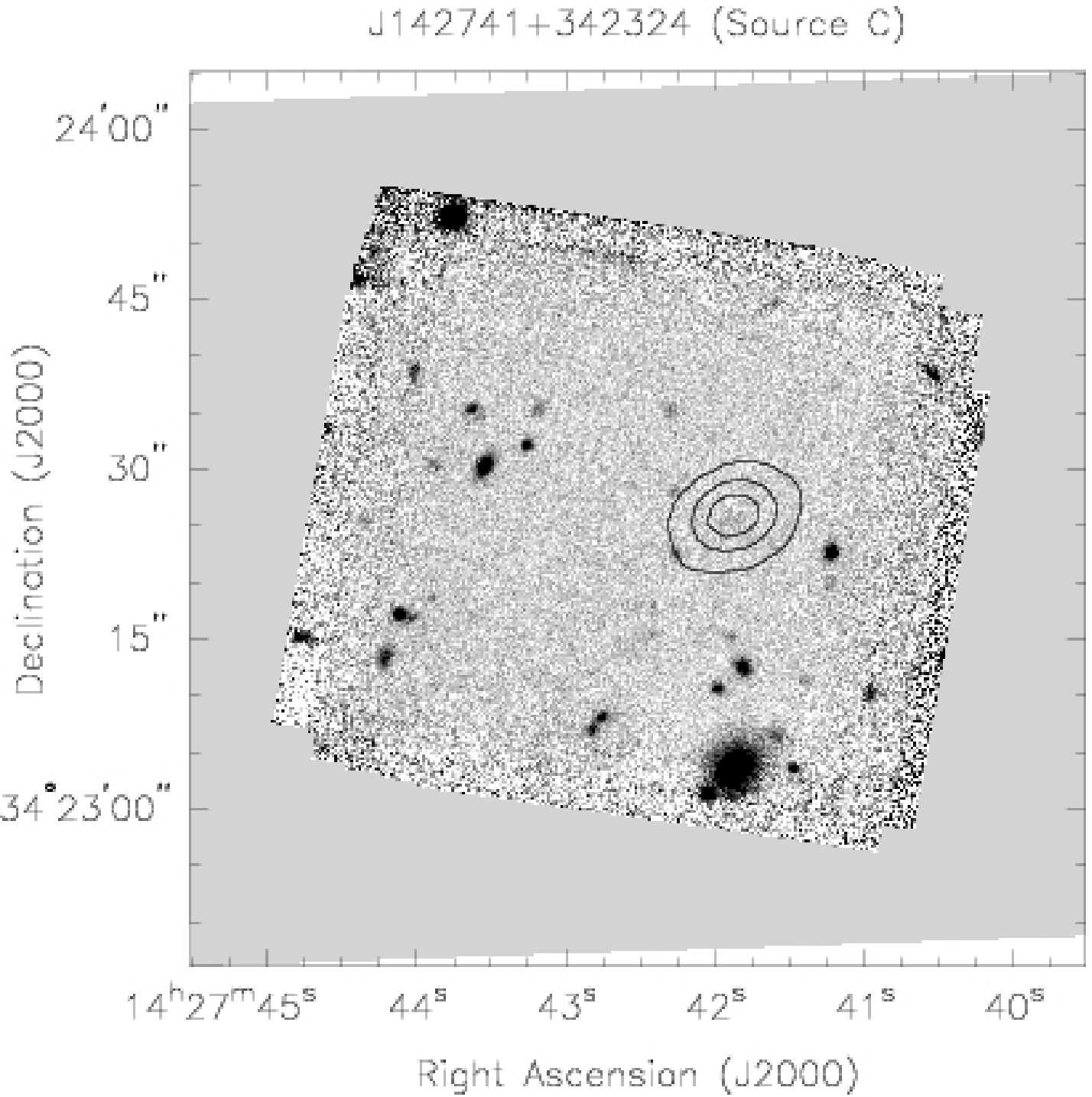}\hspace{0.03\linewidth}
\includegraphics[width=0.45\linewidth,draft=false]{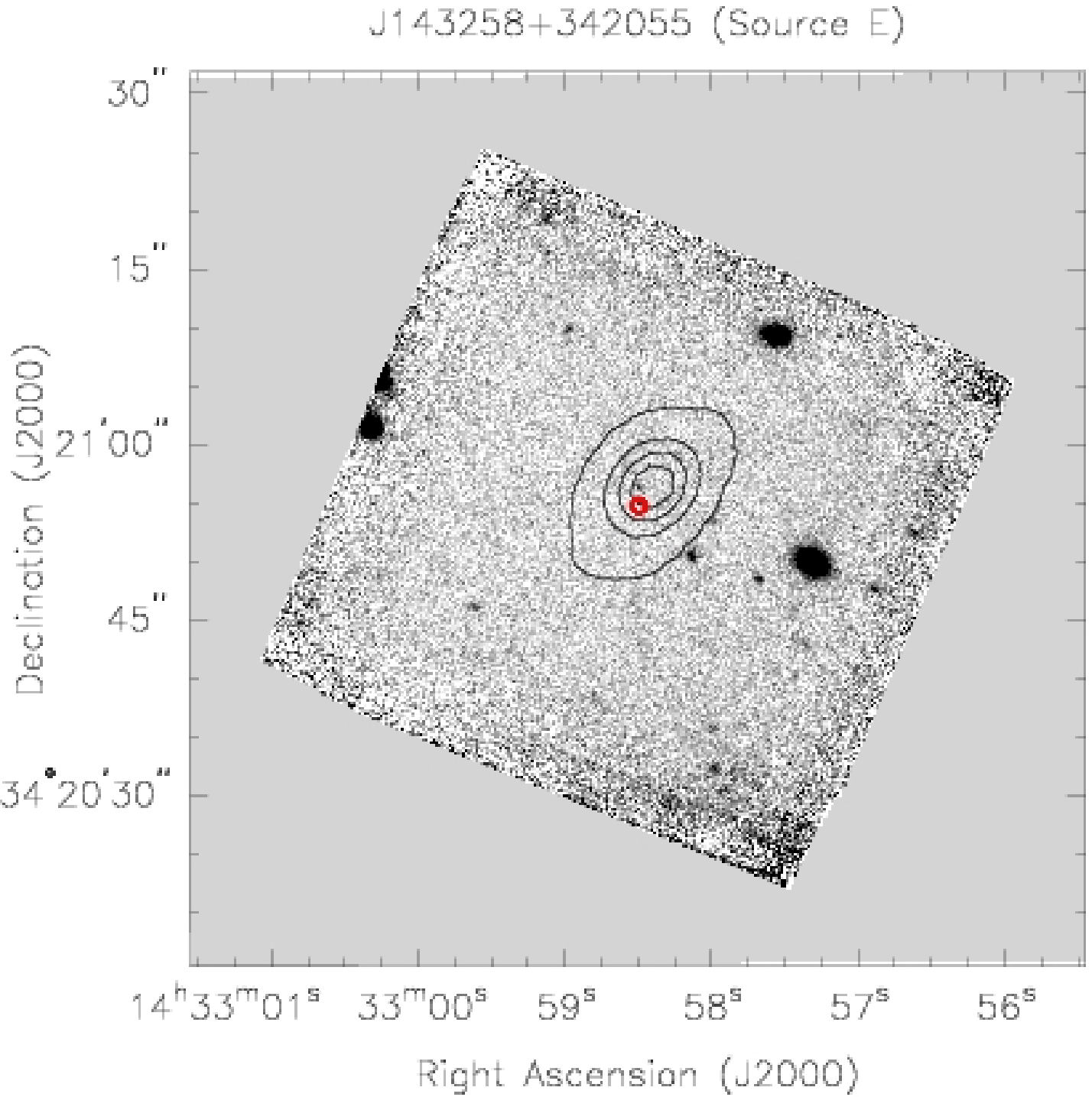}\hspace{0.03\linewidth}
\caption{\label{fig:nirccon}NIRC \ks-band images of the four \hzrg\ candidates (oriented north up). Overlaid are the 325\,MHz radio contours, from 0.5\,mJy at intervals of 1\,mJy (\srca, \srcc), from 0.3\,mJy at intervals of 0.5\,mJy (\srcb), and from 0.5\,mJy at intervals of 6.0\,mJy (\srce). To make the \ks-band ID more easily visible, the central radio contour is omitted in each map. The red circles in the images of Sources B and E mark the positions of the \xbootes\ detections (\S~\ref{sec:xray}).
}
\end{figure*}

\begin{figure*}[p]
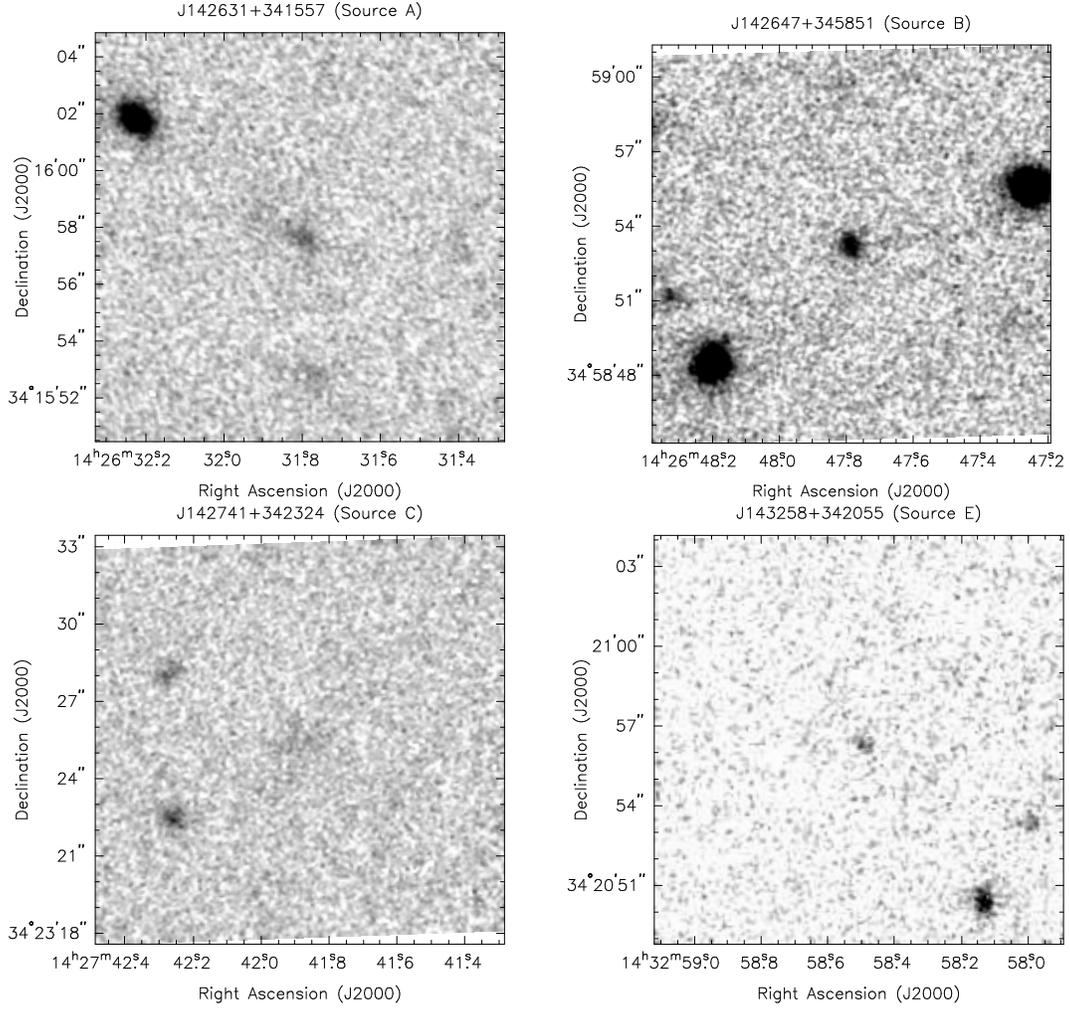

\centering
\includegraphics[width=0.4\linewidth,draft=false]{f8a.eps}\hspace{0.05\linewidth}%
\includegraphics[width=0.4\linewidth,draft=false]{f8b.eps}
\includegraphics[width=0.4\linewidth,draft=false]{f8c.eps}\hspace{0.05\linewidth}%
\includegraphics[width=0.4\linewidth,draft=false]{f8d.eps}
\caption{\label{fig:nirczoom}NIRC images of the four \hzrg\ candidates, centered on the \ks\ positions and zoomed to show the details of the radio host \ks-band morphologies. The exposure times are not the same for all four fields (Table~\ref{tab:obs}).
}
\end{figure*}

The NIRC images (Fig.~\ref{fig:nirccon}) provide much higher resolution than IRAC, and we are able to see that two of the detected sources have somewhat extended, ``fuzzy'' morphologies (Fig.~\ref{fig:nirczoom}). \srce\ is somewhat more compact, but appears to have a faint trail of material extending to the northeast. \srcb\ seems to be quite compact and relatively bright. The extended sources have \ks-band morphologies suggestive of the optical morphologies seen in \hzrgs\ such as B2\,0902+34 \citep{wvb:98}, which hints that these objects may be galaxies still in the process of formation (again arguing for a higher redshift for \srcc\ than that obtained from SED fitting). The compactness of the observed radio morphologies (unresolved in our maps, except for \srcb) suggests that we are probably seeing these radio galaxies relatively soon after the AGN turn on.  The deduced $k$-corrected rest-frame $B$-band magnitudes (Table~\ref{tab:pz}) imply optical luminosities a factor of a few times fainter than the most powerful $z > 3$ radio galaxies as studied by \citet{wvb:98} and others, and the extended emission is a factor of a few smaller (roughly $10 - 30$\,kpc as compared to $50 - 100$\,kpc for the more powerful sources). The extended radio emission of \srcb\ ($\sim 200$\,kpc if its $z_{phot}$ is correct) is typical of a classical double-lobed radio galaxy, which lends weight to the (independent) selection of a QSO2 template in the SED fitting.

Radio galaxies are among the most massive objects at all redshifts \citep{seymour:07}, but there is some evidence for a dependence of host optical luminosity on radio power \citep[\eg,][]{db:kz,mclure:04}. It is plausible that our objects are less powerful in the radio and hence might be expected to be fainter than the \kz\ relation (Fig.~\ref{fig:kz}); they might, in fact, be still forming the bulk of their stars (as suggested by their morphologies) but the AGN has already switched on. \srcb, which is more compact in the near-IR, and has more extended radio emission (suggesting that the AGN activity is less recent than in the other three sources) is slightly brighter than the \kz\ relation, again as might be expected in this scenario.

From the measured radio fluxes and spectral indices, and the best-fit photometric redshifts, we calculate rest-frame 151\,MHz luminosities (for the $z_{phot} > 3$ sources) of $\sim 1 \times 10^{27}$\,\whzsr, which implies \citep{willott:99,mccarthy:93} \lya\ luminosities $\sim 2 \times 10^{43}$\,erg\,s$^{-1}$. The corresponding \lya\ fluxes are $\sim 6 \times 10^{-17}$\,\esc, essentially independent of redshift (due to similar effects of distance on the observed \lya\ and radio fluxes, and hence the determination of rest-frame luminosities). A typical \hzrg\ \lya\ line (rest-frame FWHM $\sim 5$\,\AA) would be marginally unresolved in our spectra at $z = 0$, but beyond $z \sim 0.6$, observed flux densities would start to decrease due to broadening of the \lya\ line in the observed frame. At $z \sim 5$, the expected \lya\ flux density is $2 \times 10^{-18}$\,\esca. Since the noise in our spectra has $\sigma \sim 4 \times 10^{-19}$\,\esca, the predicted S/N is rather low ($\sim 5$), so it is perhaps unsurprising that we see no emission lines in our spectra, especially given the uncertainties ($\pm 1$\,dex) in the emission line --- radio correlation. If the line happens to fall in a region of the spectrum where noise is higher (at the position of a subtracted telluric emission line, for example), the S/N would be lower still. Dust, if present in the source, would also reduce the predicted fluxes, and hence the S/N. \srca\ has the steepest radio spectrum of our candidates, so is most similar to classical \hzrgs, and remains a good candidate for future spectroscopy. As discussed above, \srcb\ may have prominent emission lines, and spectroscopic observations would also be worthwhile.

The photometric redshifts and ``typical'' \hzrg\ morphologies support the interpretation of these objects as \hzrgs\ (and their deduced radio luminosities and $B$-band absolute magnitudes are AGN-like, as expected). At the fainter radio flux levels we probe, we seem to be seeing the more ``normal'' members of the \hzrg\ population that perhaps make up the bulk of the luminosity function at these redshifts. Indeed, comparing to the results of \citet{lo:96}, we find that our three $z > 3$ galaxies (with rest-frame $R$-band magnitudes $M_R \sim -25.5$\,mag, as computed from the photometric redshifts in Table~\ref{tab:pz}) ought to lie on the FR-I / FR-II break ($\sim 10^{27}$\,\wphz\ at these optical absolute magnitudes). 

As a sanity check, we obtain a rough estimate of the space density of these objects by calculating the comoving volume covered by the 4.9~\sqdeg\ combined radio dataset, in the redshift range 1.21 -- 4.97 (the range of photometric redshifts in Table~\ref{tab:pz}). In our chosen cosmology, this volume is $2.3 \times 10^8$\,Mpc$^3$. Since we do not claim to have exhaustively studied the radio-loud optically-faint galaxy population in this field within this redshift range, the resulting space density for our four \hzrg\ candidates of $2 \times 10^{-8}$\,Mpc$^{-3}$ should be taken as a lower limit. This compares favorably with the lower limit on radio galaxies of similar radio power ($P_{1400} > 10^{27}$\,\wphz) but steeper spectral index ($\alpha \leq -1.3$) obtained by \citet{db:06}, $1.2 \times 10^{-9}$\,Mpc$^{-3}$ at $3 < z < 4$. Since only one out of four of our \hzrg\ candidates has a spectral index as steep as this (and since both estimates are lower limits), the space density is consistent with that measured by \citeauthor{db:06}

Deeper optical and / or IR observations (particularly $J$, $H$ and / or mid-IR) would provide better constraints on our photometric redshifts. As discussed above, our objects may be only marginally too faint to be detected in our LRIS observations, so it is possible that deeper (red-sensitive) 10\,m spectroscopy could in future provide spectroscopic redshifts. Mid-IR observations would also be of interest in constraining star-formation rates, and the presence and temperature of any dust in these objects.

\acknowledgments

We thank S.~Kulkarni and collaborators for making the LRIS observations of UT 2005 June 6 (in exchange for images of GRB~0505096 on UT 2005 June $3 - 4$). We thank the MIPS \bootes\ team for MIPS data on our sources. We thank M.~Polletta and collaborators for making the SWIRE template library publicly available. Data presented herein were obtained at the W.\ M.\ Keck Observatory, which is operated as a scientific partnership among the California Institute of Technology, the University of California, and the National Aeronautics and Space Administration. The Observatory was made possible by the generous financial support of the W.\ M.\ Keck Foundation. The authors wish to recognize and acknowledge the very significant cultural role and reverence that the summit of Mauna Kea has always had within the indigenous Hawaiian community.  We are most fortunate to have the opportunity to conduct observations from this mountain. The work of SC, WvB, WdV and SAS was performed under the auspices of the U.\ S.\ Department of Energy, National Nuclear Security Administration by the University of California, Lawrence Livermore National Laboratory under contract No.\ W-7405-Eng-48. SC and WvB acknowledge support for radio galaxy studies at UC Merced, including the work reported here, with the Hubble Space Telescope and the {\em Spitzer Space Telescope} via NASA grants HST \#10127, SST \#1264353, SST \#1265551, SST \#1279182. The work of DS and PE was carried out at Jet Propulsion Laboratory, California Institute of Technology, under a contract with NASA. This work is based in part on observations made with the Spitzer Space Telescope, which is operated by the Jet Propulsion Laboratory, California Institute of Technology under a contract with NASA. Support for this work was provided by NASA through an award issued by JPL / Caltech. This work made use of images and data products provided by the NOAO Deep Wide-Field Survey, which is supported by the National Optical Astronomy Observatory (NOAO). NOAO is operated by AURA, Inc., under a cooperative agreement with the National Science Foundation. Finally, we thank the referee, R.~Windhorst, for his comments.

\end{document}